\RequirePackage{lineno}

\documentclass[12pt]{iopart}

\usepackage{preprintcover}  

\usepackage{subfigure}
\usepackage{mathrsfs}
\usepackage{graphicx}
\usepackage{multirow}
\usepackage{rotating}
\usepackage{xspace}
\usepackage{slashed}
\usepackage[hyperindex,breaklinks]{hyperref} 
\usepackage{breakurl}

\PreprintCoverPaperTitle{Search for Extra Dimensions in diphoton events from proton-proton collisions at $\sqrt{s} = 7$~TeV in the ATLAS detector at the LHC}  
\PreprintIdNumber{CERN-PH-EP-2012-289}  
\PreprintCoverAbstract{The large difference between the Planck scale and the electroweak scale, known as the hierarchy problem, is addressed in certain models through the postulate of extra spatial dimensions. A search for evidence of extra spatial dimensions in the diphoton channel has been performed using the full set of proton-proton collisions at $\sqrt{s} = 7$~TeV recorded in 2011 with the ATLAS detector at the CERN Large Hadron Collider.
This dataset corresponds to an integrated luminosity of $4.9$~fb$^{-1}$. The diphoton invariant mass spectrum is observed to be in good agreement with the Standard Model expectation. In the context of the model proposed by Arkani-Hamed, Dimopoulos and Dvali, $95\, \%$ confidence level lower limits of between 2.52~TeV and 3.92~TeV are set on the ultraviolet cutoff scale~$M_S$ depending on the number of extra dimensions and the theoretical formalism used.
In the context of the Randall--Sundrum model, a lower limit of 2.06 (1.00)~TeV at 95\,\% confidence level is set on the mass of the lightest graviton for couplings of $\coupling$ = 0.1 (0.01). Combining with the ATLAS dilepton searches based on the 2011 data, the 95\,\% confidence level lower limit on the Randall--Sundrum graviton mass is further tightened to 2.23 (1.03)~TeV for $\coupling$ = 0.1 (0.01).} 
\PreprintJournalName{NJP}  

\def\EtIso{\ensuremath{E_{\rm T}^{\rm iso}}}
\def\EtGamma{\ensuremath{E_{\rm T}^{\gamma}}}
\def\Pythia{{\sc Pythia}}
\def\Diphox{{\sc Diphox}}
\def\Sherpa{{\sc Sherpa}}
\def\GeV{${\rm GeV}$}

\def\Ggg{\ensuremath{G \rightarrow \gamma \gamma}}
\def\coupling{\ensuremath{k/\overline{M}_{Pl}}}
\def\mpl{\ensuremath{\overline{M}_{Pl}}}

\begin{document}

\title[Search for Extra Dimensions in diphoton events with the ATLAS detector]{Search for Extra Dimensions in diphoton events from proton-proton collisions at $\sqrt{s} = 7$~TeV in the ATLAS detector at the LHC}
\author{The ATLAS Collaboration}

\begin{abstract}
The large difference between the Planck scale and the electroweak scale, known as the hierarchy problem, is addressed in certain models through the postulate of extra spatial dimensions. A search for evidence of extra spatial dimensions in the diphoton channel has been performed using the full set of proton-proton collisions at $\sqrt{s} = 7$~TeV recorded in 2011 with the ATLAS detector at the CERN Large Hadron Collider.
This dataset corresponds to an integrated luminosity of $4.9$~fb$^{-1}$. The diphoton invariant mass spectrum is observed to be in good agreement with the Standard Model expectation. In the context of the model proposed by Arkani-Hamed, Dimopoulos and Dvali, $95\, \%$ confidence level lower limits of between 2.52~TeV and 3.92~TeV are set on the ultraviolet cutoff scale~$M_S$ depending on the number of extra dimensions and the theoretical formalism used.
In the context of the Randall--Sundrum model, a lower limit of 2.06 (1.00)~TeV at 95\,\% confidence level is set on the mass of the lightest graviton for couplings of $\coupling$ = 0.1 (0.01). Combining with the ATLAS dilepton searches based on the 2011 data, the 95\,\% confidence level lower limit on the Randall--Sundrum graviton mass is further tightened to 2.23 (1.03)~TeV for $\coupling$ = 0.1 (0.01).
\end{abstract}

\submitto{\NJP}
\maketitle

\section{Introduction}

One of the major goals of particle physicists in the past decades has been to solve the  hierarchy problem: 
the fact that the electroweak scale is 16 orders of magnitude smaller than 
the Planck scale. One possible solution to the hierarchy problem is the existence of extra spatial dimensions. In this paradigm, gravity appears much weaker than the other interactions because it is diluted by the presence of extra spatial dimensions.

In the Randall-Sundrum (RS) model~\cite{RSmodel}, a five-dimensional geometry is assumed, in which the fifth
dimension is compactified with length  $r_c$. There are two four-dimensional branes sitting in a five-dimensional bulk with a
``warped'' geometry. The Standard Model (SM) fields are located on the so-called  TeV~brane, while gravity
originates from the other brane, called the Planck~brane. Gravitons are capable of propagating in the bulk.
 Mass scales on the TeV~brane correspond to mass scales on the Planck brane ($M_{Pl}$) as given by $M_D = M_{Pl} e^{-k\pi r_c}$, where k is the curvature scale of the extra dimension.
The observed hierarchy of scales can be naturally reproduced
in this model if $kr_c \approx 12$~\cite{coupling}. 
The compactification of the extra dimension gives rise to a Kaluza--Klein~(KK) tower of graviton excitations~$G$, a set of four-dimensional particles with increasing masses.
The phenomenology can be described in terms of the mass of the lightest KK graviton excitation ($m_{G}$) and the dimensionless coupling
to the SM fields, \coupling, where $\mpl = M_{Pl}/\sqrt{8\pi}$ is the
reduced Planck scale. From theoretical arguments~\cite{coupling},
\coupling \ values in the range [0.01, 0.1] are preferred. The lightest
RS~graviton is expected to be a fairly narrow resonance for $\coupling
< 0.3$. The most stringent experimental limits on RS gravitons have
been obtained at the LHC: the current best $95\, \%$ confidence level (CL) lower
limits on the graviton mass for  \coupling = 0.1 are
2.16~TeV~\cite{dilepton} and 2.14~TeV~\cite{Chatrchyan2012158}. The former
limit has been obtained from an analysis of the $G\rightarrow
ee/\mu\mu$ channels in data corresponding to an
integrated luminosity of 5~fb$^{-1}$ collected by the ATLAS
experiment. The latter limit has been obtained from an analysis of the
same channels in data corresponding to an
integrated luminosity of $\approx$5~fb$^{-1}$ collected by the CMS experiment.
The limits from earlier searches at the Tevatron can be found in Refs.~\cite{D0RS,CDFRS}. 

Arkani-Hamed, Dimopoulos, and Dvali (ADD)~\cite{ADD}  proposed a different scenario for extra dimensions. Motivated by the weakness of gravity, 
they postulated the existence of $n$ flat additional spatial dimensions compactified with radius $R$, and proposed a model in which only gravity propagates in the extra dimensions. The fundamental Planck scale in the $(4 + n)$-dimensional spacetime, $M_{D}$, is related to the apparent scale \mpl\ by Gauss's law: $\mpl^{2} = M_{D}^{n+2}R^{n}$. 
The mass splitting of the graviton KK modes is $1/R$ for each of the $n$ extra dimensions.
In the ADD model, resolving the hierarchy problem usually requires small values of $1/R$, giving rise to an
almost continuous spectrum of KK graviton states. The symbol~$G$ is used to denote KK~graviton states in both the RS and ADD models.
The existence of ADD extra dimensions can manifest itself in proton-proton ($pp$) collisions through a variety of processes, including virtual graviton 
exchange as well as direct graviton production.
While processes involving direct graviton emission depend on $M_D$, effects involving virtual gravitons depend on the ultraviolet cutoff of the KK~spectrum, denoted $M_{S}$.
The strength of gravity in the presence of extra dimensions is typically parameterised 
by  $\eta_{G} = F/M_S^4$, where
$F$ is a dimensionless parameter of order unity reflecting the dependence of virtual 
KK graviton exchange on  the number of extra dimensions. Several theoretical 
formalisms exist in the literature, using different definitions 
of $F$ and, consequently, of $M_{S}$:

\begin{equation}
\label{eq:ADD_GRW}
F  =  1 ,    \textrm{ (GRW)~\cite{GRW}}; \\
\end{equation}

\begin{equation}
F = \left\{ \begin{array}{ll}
\textrm{log}\left( M_{S}^{2}/\hat s\right) & \textrm{$n = 2$} \\
2/(n-2) & \textrm{$n > 2$}
\end{array} \right.
, \textrm{ (HLZ)~\cite{HLZ};} \\
\label{eq:ADD_HLZ}
\end{equation}

\begin{equation}
\label{eq:ADD_Hew}
F  =   \pm2/\pi ,    \textrm{ (Hewett)~\cite{Hewett}};
\end{equation}

\noindent where $\sqrt{\hat s}$
is the centre-of-mass energy of the parton-parton collision.
The effect of ADD graviton exchange manifests itself as a non-resonant deviation from the SM background expectation.
The effective diphoton cross section is the result of the SM and ADD amplitudes, as well as their interference. The interference term in the effective cross section is linear in $\eta_{G}$ and the pure graviton exchange term is quadratic in $\eta_{G}$.
In previous analyses by the ATLAS and CMS collaborations, $95\, \%$ CL lower limits on $M_{S}$ of 2.26--3.52 TeV~\cite{ATLAS:2011ab} and 2.4--3.8~TeV~\cite{CMSggnew} respectively were set, depending on the theoretical formalism used. Searches for ADD virtual graviton effects have also been performed at other colliders, {\em e.g.} at HERA~\cite{H1ADD,ZEUSADD}, LEP~\cite{ALEPHADD,DELPHIADD,LEPNOTEADD} and the Tevatron~\cite{TevatronADD,D0ADDjets}.

This Letter reports the results of a search for extra dimensions in the diphoton channel using the full data sample 
recorded by the ATLAS detector in 2011, corresponding to a total integrated luminosity
of $4.9$~fb$^{-1}$ of $pp$ collisions at $\sqrt s= 7$~TeV. The results are interpreted in the context of both the ADD and RS scenarios. The dilepton and diphoton channels are particularly sensitive for this search due to the clean experimental signature, excellent mass resolution and modest backgrounds. The branching ratio for graviton decay to two photons is twice the value of that for graviton decay to any individual charged-lepton pair.
This search, based on the full $\sqrt s = 7$~TeV $4.9$~fb$^{-1}$ dataset, is an extension of an earlier study~\cite{ATLAS:2011ab} performed using $1.0$~to~$2.1$~fb$^{-1}$ of data. 
The earlier data are included in the present study but they benefit from improvements in the reconstruction and calibration procedures. Moreover, the techniques used for the
background estimation and their associated uncertainties have been also significantly improved. This results to an overall improvement of a factor 2 to 3 in the excluded production
cross section of diphoton events predicted by the Randall-Sundrum model, corresponding to a graviton mass limit which is 10\% to 15\% larger.
\section{The ATLAS detector}

The ATLAS detector is a multipurpose particle
physics instrument with a forward-backward symmetric cylindrical
geometry and near $4\pi$ solid angle coverage.\footnote{ATLAS uses a right-handed coordinate system with its origin at the nominal interaction point (IP) in the centre of the detector and the $z$-axis along the beam pipe. The $x$-axis points from the IP to the centre of the LHC ring, and the $y$ axis points upward. Cylindrical coordinates $(r,\phi)$ are used in the transverse plane, $\phi$ being the azimuthal angle around the beam pipe.} 
A detailed description of the ATLAS detector can be found in Ref.~\cite{DetectorPaper}.
The inner tracking detector~(ID)  consists of a silicon pixel
detector, a silicon microstrip detector, and a transition
radiation tracker. The ID is surrounded
by a superconducting solenoid that provides a 2~T axial magnetic field. The ID allows an accurate reconstruction
of tracks from the primary $pp$
collision and also identifies tracks from secondary
vertices, permitting the efficient identification of photon
conversions.

The electromagnetic calorimeter (ECAL) is a lead/liquid-argon (LAr) sampling calorimeter with an accordion
geometry. It is divided into a barrel section, covering the
pseudorapidity\footnote{The pseudorapidity is defined in terms of the polar angle $\theta$ as $\eta=-\ln\tan(\theta/2)$.} region $\left|\eta\right| < 1.475$, and two endcap sections,
covering the pseudorapidity regions $1.375 < \left|\eta\right| < 3.2$. It consists of three longitudinal layers for $\left|\eta\right |< 2.5$ and two for $2.5 < \left|\eta\right| < 3.2$. 
Up to $\left|\eta\right| < 2.4$, the first one uses highly granular ``strips" segmented in the $\eta$~direction for efficient event-by-event discrimination between single photon showers and two overlapping showers originating from $\pi^0$ decay.  The second layer collects most of the energy deposited in the electromagnetic calorimeter by photon showers.
Significant energy deposits in the third layer are an indication for leakage beyond the ECAL from a high energy shower. The measurements from the third layer are used to correct for this effect.
A thin presampler layer in front of the accordion calorimeter, covering the pseudorapidity
interval $\left|\eta\right| < 1.8$, is used to correct for energy loss
upstream of the calorimeter.

The hadronic calorimeter (HCAL), surrounding the
ECAL, includes a central ($\left|\eta\right| < 1.7$) iron/scintillator
tile calorimeter, two endcap ($1.5 < \left|\eta\right| < 3.2$) copper/LAr calorimeters and two forward calorimeters that extend the coverage to $\left|\eta\right| < 4.9$, using copper and tungsten as absorber. The muon spectrometer, located beyond the calorimeters, consists
of three large air-core superconducting toroid systems, instrumented with precision tracking chambers as well as fast detectors for triggering.

\section{Trigger and event selection}

The analysis uses data recorded by ATLAS between March and October
2011 during stable-beam periods of $pp$ collisions at $\sqrt{s} = 7$~TeV.
A three-level trigger system is used to select events containing two photon candidates. The first level trigger is hardware-based and 
relies on a readout with coarse granularity. The second- and third-level triggers, collectively
referred to as the high-level trigger (HLT), are implemented in
software and exploit the full granularity and energy calibration
of the calorimeter.
Selected events have to satisfy a diphoton trigger where each photon is required to satisfy, at the HLT level, a transverse energy requirement $\EtGamma>$ 20~\GeV\ and a set of requirements~\cite{ATLASphotons,ATLASphotonid} on the shape of the energy deposit.
This set includes criteria for the energy leakage into the hadronic calorimeter and for the width of the shower in the second layer of the ECAL. 
This trigger is nearly $100 \, \%$ efficient for diphoton events passing the final selection requirements.

Only events satisfying the standard ATLAS data quality requirements and having at least one primary collision vertex with at least three 
tracks are kept.  Photon candidates reconstructed in the precision regions of the barrel ($\left|\eta\right| < 1.37$) or the 
endcaps ($1.52 < \left|\eta\right| < 2.37$), with $\EtGamma>$ 25~\GeV\ and satisfying the same requirements on the shower shape as the trigger are preselected. These requirements on the photon candidates are referred to as the loose selection. The two highest-\EtGamma\ photon candidates each have to satisfy a set of stricter requirements, referred to as the tight photon definition, which includes a more stringent selection on the shower width in the second layer of the ECAL and additional requirements on the energy distribution in the first layer of the ECAL. 
In addition, the two photon candidates of interest have to satisfy a calorimetric isolation requirement, $\EtIso<5$~\GeV. This isolation is computed by summing the transverse energy over the cells of the calorimeters in a cone of radius $\Delta R=\sqrt{(\eta-\eta^{\gamma})^2+(\phi-\phi^{\gamma})^2}<0.4$ surrounding the photon candidate. Here $(\eta, \phi)$ denotes the cell position and $(\eta^{\gamma},\phi^{\gamma})$ denotes the position of the photon cluster~\cite{Lampl:1099735}. 
The transverse energy contribution in the centre of the cone, which
contains the photon cluster, is subtracted. Corrections, based on simulations, are applied to account for the expected energy leakage from the photon into the isolation region.
An ambient-energy correction, based on the measurement of low-transverse-momentum jets~\cite{ambient} is also applied, on an event-by-event basis, to remove the contributions from the underlying event and from pileup which results from the presence of multiple $pp$~collisions within the same or nearby bunch crossings.

\section{Event simulation}

Monte Carlo (MC) simulations are used to study the response of the detector to various RS and ADD scenarios as well as SM diphoton processes. 
The MC generators used in this analysis are listed below. For samples generated using \Pythia~\cite{pythia}, the ATLAS parameter tunes~\cite{ATLAStune} 
are used in the event generation. For samples generated using \Pythia\ or \Sherpa~\cite{sherpa},
the generated events are processed through the {\sc
  Geant4}~\cite{geant4} detector simulation~\cite{detsimu}. Then the
simulated events are reconstructed with the same procedure as is used
for the data. To reproduce the pileup conditions of the data, the
simulated events are reweighted to match the observed distribution of
the average number of interactions per bunch crossing. The determination of the number of interactions per bunch crossing is discussed in Ref.~\cite{NBinteractions}.

\Pythia\ 6.424 with MRST2007LOMOD~\cite{mrst2007lomod} parton distribution functions (PDFs) is used to simulate the SM diphoton production at leading order (LO). In addition, the next-to-leading (NLO) correction and the contribution of the fragmentation processes are evaluated using \Diphox~\cite{diphox} 1.3.2 with MSTW2008NLO~\cite{mstw2008nlo} PDFs and lead to an invariant-mass-dependent correction of the \Pythia\ prediction. The so-called box contribution $gg\rightarrow\gamma\gamma$ through a quark loop is included in both the \Pythia\ and the \Diphox\ predictions. From the point of view of power counting, this diagram is a next-to-next-to-leading (NNLO) contribution, but the gluon luminosity at the LHC is so large compared to the quark-antiquark one that this process represents a contribution that is comparable to the $q\bar{q}\rightarrow\gamma\gamma$ process. Samples of RS events are produced using the same generator and PDFs as for the SM diphoton samples. They are used to study the selection efficiency and the shape of the reconstructed invariant mass spectrum for various values of the graviton mass $m_G$ between 400~\GeV\ and 3000~\GeV\ and values of \coupling\ in the range $[0.01,0.1]$.
ADD models are simulated for various $M_{S}$ values using \Sherpa\ 1.3.1 with CTEQ6L~\cite{cteq6l} PDFs. The ADD MC samples are used to evaluate the number of signal events passing the selection as a function of $M_{S}$. An NLO K-factor of 1.75$\pm$0.10 is considered for the RS scenario and 1.70$\pm$0.10 is considered for the ADD scenario.
These K-factors have been provided by the authors of Refs.~\cite{kRS,kADD}. 

\section{Background estimate}

The dominant background in this analysis is the irreducible background due to SM $\gamma \gamma$ production.
Another significant background component is the reducible background which arises from events in which one or 
both of the photon candidates result from a different object being misidentified as a photon. This background 
is dominated by $\gamma + {\rm jet}$~($j$) and $j+j$ events, with one or both photon candidates arising from jets. 
Backgrounds with electrons faking photons, such as the Drell--Yan production of electron-positron pairs as well as 
$W/Z + \gamma$ and $t\overline{t}$ processes, have been verified to be negligible after the event selection.

We first describe the tools and samples that are used to predict the contributions from the SM $\gamma \gamma$ background to the high-mass signal region, as well as their main limitation.
We then describe in detail the two-step procedure that is used to overcome this limitation and to obtain our background estimate.

The estimate of the irreducible background makes use of the NLO calculations~\cite{diphox} that are available for the SM $\gamma \gamma$ 
production processes. The uncertainty in the absolute normalisation of the $\gamma \gamma$ cross section is, however,
quite substantial: $\simeq 20 \,\%$ ($\simeq 25\, \%$) uncertainty on the cross section for $\gamma \gamma$ events with an invariant mass
$m_{\gamma \gamma} > 200$~GeV ($m_{\gamma \gamma}>1200$~GeV) with both photons in the detector acceptance. 
The background estimates, which are discussed in detail below,  are
therefore normalised to the data in the low-mass control
region\footnote{The exact values of the upper and lower bound of the
  low-mass control region have been chosen to coincide with the
  boundaries of the first bin in Table~\ref{tab:bkgestimation}. The
  results of this search are not altered by small variations of the
  control region edges nor a change of the binning.} ($[142,409]$~GeV)
where the presence of any signal beyond the SM has been excluded by previous searches. Compared to the use of NLO calculations to predict the absolute rate of SM $\gamma \gamma$ 
production in the high-mass signal region, the use of NLO calculations for the extrapolation of the SM $\gamma \gamma$ 
production rates from the low-mass control region to the high-mass signal region results in significantly smaller uncertainties.
The extrapolation method is described in detail below, along with quantitative estimates of the final uncertainties.

 Following these considerations, the estimate the $m_{\gamma \gamma}$~spectrum of the background events is done in two steps. First the shape of the $m_{\gamma \gamma}$~spectrum is determined, separately for each background component. Then the normalisation for each component is determined using the low-mass control region.

The shape of the $m_{\gamma \gamma}$ spectrum from SM $\gamma \gamma$ production is
estimated using simulated events, reweighting the \Pythia\ samples to the differential (in $m_{\gamma \gamma}$) cross 
section predicted using \Diphox. 

The shape of the reducible background is estimated using data-driven techniques. It is split into three components: $\gamma + j$ events (the leading-$E_{\rm T}$ photon candidate is due to a real photon),  $j + \gamma$ (fake leading photon candidate) and $j+j$. The shapes of the $j+ \gamma$ and $\gamma + j$ differ significantly because of the strong $p_T$-dependence of the probability to mis-identify a jet as a photon.
Separately for each component, several data control samples that are enriched in the given component are defined. 
The shape of the $\gamma + j$ background is estimated from a data control sample
that is selected in the same way as the signal sample, except that the subleading $\gamma$~candidate is required to pass
loose identification criteria and to fail tight identification criteria. The control samples used to extract the shapes of the 
$m_{\gamma \gamma}$ spectra of $j + \gamma$ and $j + j$ backgrounds are defined following the same approach,
{\em i.e.} by requiring the leading photon candidate ($j + \gamma$ background) or both candidates ($j + j$ background) to pass loose and to fail tight identification criteria.
Since the data control samples contain relatively few events at high $m_{\gamma \gamma}$, a fit to a smooth function 
of the form $f(m_{\gamma \gamma}) = p_1 \times \left( m_{\gamma \gamma} \right) ^{p_2 + p_3 \log m_{\gamma \gamma}}$, where the $p_i$ are free parameters, 
is used to extrapolate the reducible background shapes to higher masses. This functional form has been used in the CDF dijet search~\cite{cdfdijet} and in previous ATLAS searches in the dilepton~\cite{ATLASdilep}, dijet~\cite{ATLASdijet}, photon--jet~\cite{ATLASphotjet} and diphoton~\cite{ATLAS:2011ab}  channels, and it describes the shapes of the data control samples well. In contrast to the earlier 
version of this analysis~\cite{ATLAS:2011ab}, the shapes of the three components of the reducible background are modelled 
separately.

\begin{figure}
\begin{center}
\includegraphics[width=\textwidth]{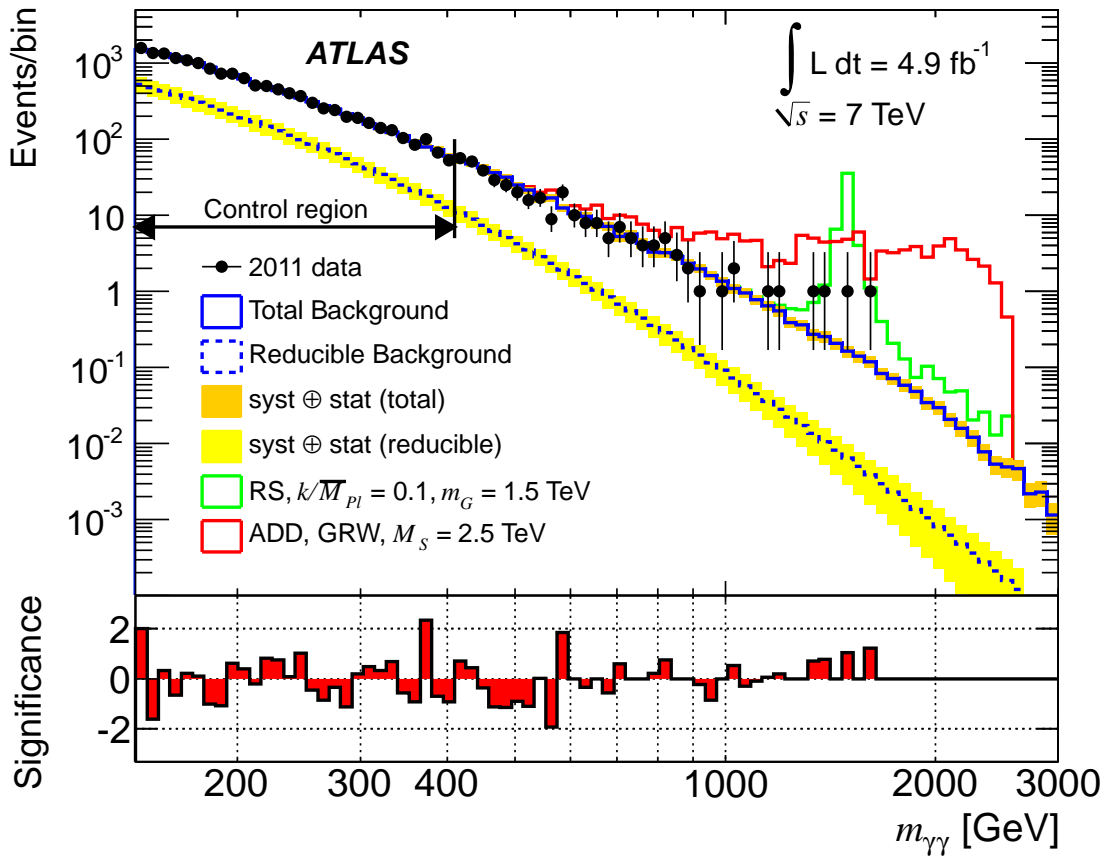}
\caption{Observed invariant mass distribution of diphoton
  events. Superimposed are the SM background expectation and the
  expected signals for an example each for RS and ADD models. In addition
  to the total background, the contribution from the reducible
  component is shown. The black arrow indicates the control region. To compensate the rapid decrease of the
  spectrum, the bins have been chosen to have constant logarithmic
  width. Specifically, the ratio of the upper to lower bin boundary is
  equal to 1.038 for all bins, and the first bin starts at
  142~GeV. The bin-by-bin significance of the difference between
  data and predicted background is shown below the main
  plot. Following the convention of Ref.~\cite{significance}, the
  significance is set to zero for bins with insignificant deviations with respect to a small expected
  background. This concerns mainly the bins at large $m_{\gamma\gamma}$
  where a fraction of an event is expected (on average) and where zero
  events are observed.}
\label{fig:databkg}
\end{center}
\end{figure}

\begin{figure}[h!]
\begin{center}
\includegraphics[width=\textwidth]{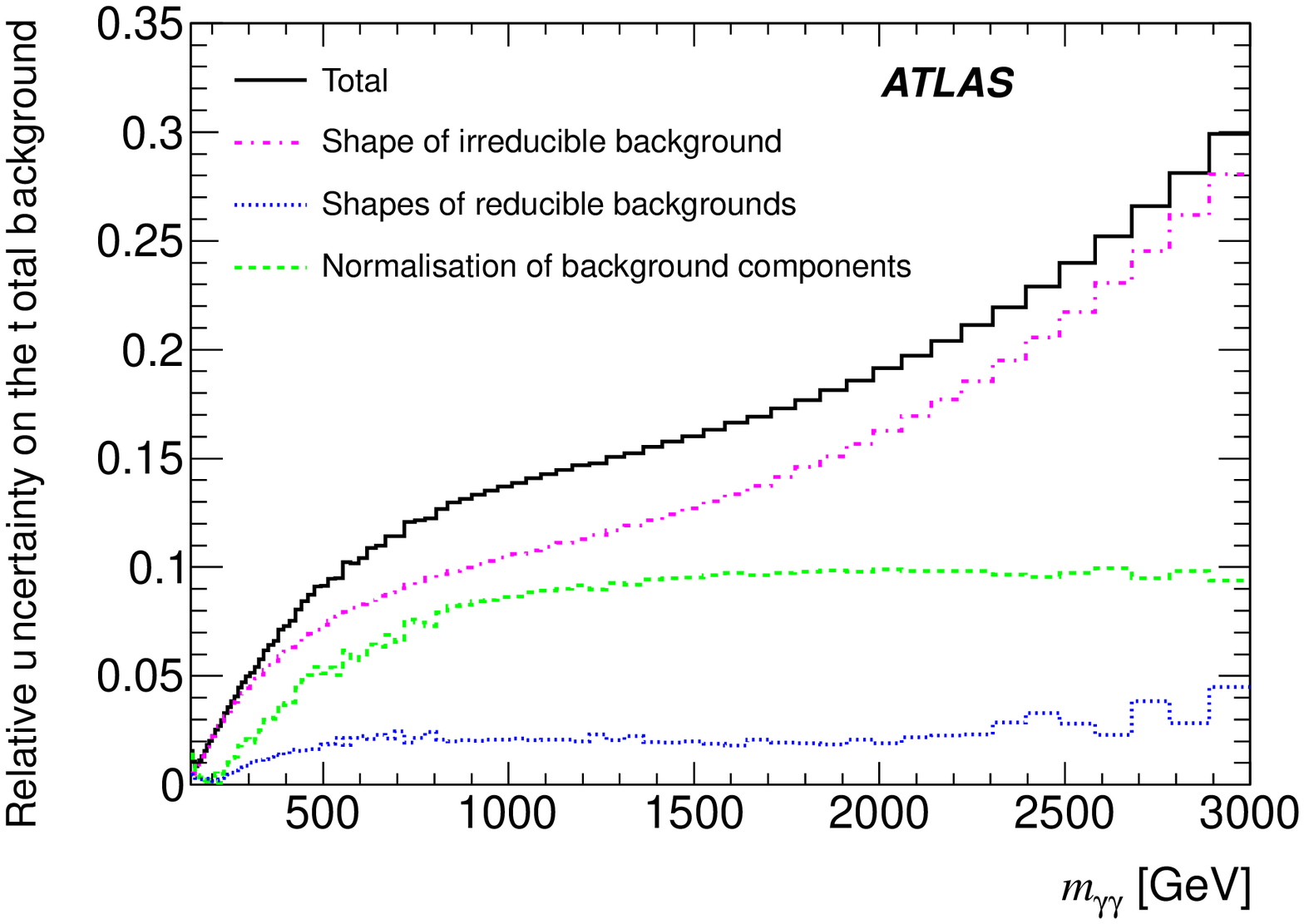}
\caption{Uncertainty in the total background expectation as a function of $m_{\gamma \gamma}$. 
Also shown are the individual contributions from the uncertainties in the shapes of the irreducible and the reducible components, 
as well as the contribution from the uncertainties in the yields (normalisations) of the different background components 
extracted from the low-mass control region. Since the number of events
in the low-mass control region is dominated by the lowest masses, the
background uncertainties are particularly small at the low end of the
control region.}
\label{fig:bkgndtotalsystematic}
\end{center}
\end{figure}

To determine the contributions of each of the four sources (the
irreducible $\gamma\gamma$ component plus the three reducible
components, $\gamma + j$, $j + \gamma$ and $j+j$) to the low-mass control region, a two-dimensional
template fit to the distributions of the calorimetric isolation (\EtIso) of the two photon candidates is used. 
For the purpose of this fit, the isolation requirement in the event selection is relaxed to $\EtIso<25$~\GeV.
This method has been used previously in Refs.~\cite{ATLASsm,ATLAShiggs} and in an earlier version of this analysis~\cite{ATLAS:2011ab}. Templates for the \EtIso\ distribution of true photons and of fake photons from jets are both determined 
from data. The shape for fake photons is determined using a sample of photon candidates that fail the tight requirement but pass a set of requirements referred as the non-tight selection. This non-tight selection includes the same requirements as the loose selection plus additional requirements that reduce the correlation between the identification and the isolation. The shape for true photons is found from the sample of tight photon candidates, after subtracting the fake photon shape
normalised to match the number of candidates with large values of \EtIso\ ($\EtIso > 10$~GeV; this control region is dominated by fake photons). Both the signal templates and fake templates are constructed separately for leading and subleading photon candidates. The significant correlation ($\sim 20 \, \%$) between the \EtIso\ values of the two fake photons in the $j+j$ background is included in the two-dimensional template for this background component. 
The background expectation as a function of $m_{\gamma \gamma}$ is summarised in Fig.~\ref{fig:databkg} and Table~\ref{tab:bkgestimation}. Fig.~\ref{fig:bkgndtotalsystematic} details the different contributions to the uncertainty in the background expectation. 

\begin{table*}[tb]
\caption{Expected numbers of events from the irreducible and reducible
background components, as well as the total background prediction and observed
numbers of events in $m_{\gamma \gamma}$ bins. The table quotes the
sum of the statistical and systematic uncertainties in the background expectations. The boundaries of all bins in this table have been chosen to coincide with bin 
boundaries in Fig.~\ref{fig:databkg}. As discussed in the text, the background estimate is normalised to the data in the
low-mass control region $[142,409]$~GeV. Since the uncertainties in the normalisation are strongly anticorrelated between the reducible 
and irreducible components, their impact on the estimate of the total background is smaller than
their impact on the individual components. By construction, the uncertainties in the control region are 100\% anticorrelated and the total background expectation is identical to the number of observed events.}
\begin{center}
\begin{tabular}{  c |c c c | c} \hline
\hline
Mass Window  & \multicolumn{3}{c|}{Background Expectation}  & Observed \\
(GeV)        & Irreducible    & Reducible   & Total         & Events   \\
\hline
\hline
$[142,409]$ & \multirow{2}{*}{$10195\pm1092$} & \multirow{2}{*}{$4586\pm1092$} & \multirow{2}{*}{$14781$} & \multirow{2}{*}{$14781$} \\
Control region  & & & &\\
\hline
$[409,512]$ & $192\pm26$ & $43\pm10$ & $235\pm20$ & $221$ \\
$[512,596]$ & $57\pm8$ & $10.7\pm2.7$ & $68\pm7$ & $62$ \\
$[596,719]$ & $35\pm5$ & $5.4\pm1.5$ & $40\pm4$ & $38$ \\
$[719,805]$ & $12.0\pm1.8$ & $1.4\pm0.4$ & $13.3\pm1.6$ & $13$ \\
$[805,901]$ & $7.8\pm1.2$ & $0.70\pm0.23$ & $8.5\pm1.1$ & $10$ \\
$[901,1008]$ & $4.6\pm0.7$ & $0.35\pm0.13$ & $5.0\pm0.7$ & $2$  \\
$[1008,1129]$ & $2.7\pm0.4$ & $0.18\pm0.07$ & $2.8\pm0.4$ & $2$ \\
$[1129,1217]$ & $1.14\pm0.18$ & $0.064\pm0.028$ & $1.21\pm0.18$ & $2$ \\
$[1217,1312]$ & $0.72\pm0.12$ & $0.040\pm0.018$ & $0.76\pm0.11$ & $0$ \\
$[1312,1414]$ & $0.50\pm0.08$ & $0.024\pm0.012$ & $0.53\pm0.08$ & $2$ \\
$[1414,1644]$ & $0.61\pm0.10$ & $0.024\pm0.013$ & $0.63\pm0.10$ & $2$ \\
$[1644,2889]$ & $0.39\pm0.08$ & $0.013\pm0.009$ & $0.40\pm0.07$ & $0$ \\
\hline 
\hline
\end{tabular}
\label{tab:bkgestimation}
\end{center}
\end{table*}

The uncertainties in the shape of the irreducible background are
dominated by the uncertainties in the PDFs that are used with
\Diphox. Those uncertainties are evaluated using the MSTW2008 NLO
eigenvector PDF sets and using CTEQ6.6 and MRST2007LOMOD PDF sets for
comparisons. The spread of the variations includes the difference
between the central values obtained with the different PDF sets. Smaller contributions
arise from residual imperfections in the simulation of the isolation
variable~\EtIso\ and from higher-order contributions that are not
included in the \Diphox\ model. The latter contribution is evaluated by varying in a coherent and an incoherent way the renormalisation, the initial factorisation and the final factorisation scales by a factor two around their nominal value, which is the invariant mass of the diphoton system. The shape predictions from \Diphox\ are found to be in good agreement with those obtained using an alternative NLO generator, {\sc MCFM}~\cite{MCFM} (with the same PDF set as \Diphox). The small difference between the two generators is assigned as an additional systematic uncertainty. It represents a negligible contribution for $m_{\gamma \gamma} > 1$~TeV. The uncertainties in the shape 
of the reducible background arise from the finite size of the background control samples, and from the extrapolation 
of the background shapes from the control sample to the signal region. The latter are assessed by varying the definition of the
loose selection requirement which is used to define the control samples. The shape uncertainty of the reducible background has a small impact on the total background (Fig.~\ref{fig:bkgndtotalsystematic}). This is mainly because the reducible background is a small contribution to the total (Table~\ref{tab:bkgestimation}). The uncertainties in the normalisation of the background components are dominated by the uncertainties in the templates used in the fit to the \EtIso\ distributions. The uncertainties in the templates are assessed by varying the definition of the non-tight selection requirement used in the extraction of the templates. They are then propagated to the normalisations by repeating the template fits with the varied templates.

\section{Systematic uncertainties in the signal models}

Experimental systematic uncertainties on the signal yields are
evaluated for both the RS and the ADD signals. The main systematic
uncertainties arise from the non-perfect simulation of the quantities used for the identification of photons and in the isolation requirements. 
For the former, the data-driven estimates from Ref.~\cite{ATLASphotonid} are used, together with an extrapolation to larger values of $p_T$, which are relevant in the present search. In the RS scenario, these uncertainties on the photon identification translate into a relative uncertainty of 4.7\% to 3.3\% on the signal yield for $m_{G}$ values of 0.5~TeV to 2.5~TeV. In the ADD scenario, the relative uncertainty on the signal yield is 4.5\% with a weak dependence on $M_{S}$. 
For the latter, a study comparing the data sample and simulation has shown that a 1~GeV shift  of the isolation distribution in the simulation covers possible discrepancies between data and simulation in the signal region. In the RS scenario, these uncertainties translate into a relative uncertainty of 4\% to 7\% on the signal yield for $m_G$ values of 0.5~TeV to 2.5~TeV. In the ADD scenario the relative uncertainty on the signal yield is of $\sim 6\%$ with a weak dependence on $M_{S}$.  The uncertainty on the integrated luminosity for the complete 2011 dataset is 3.9\%, based on the calibration described in Refs.~\cite{NBinteractions,ATLASlumi2011} plus an additional uncertainty for the extrapolation to the later data-taking period with higher instantaneous luminosity. An uncertainty of $2\,\%$ on the trigger efficiency arises from differences between the measured trigger efficiency in data and simulated samples. Given that the sum in quadrature of all experimental systematic uncertainties depends only weakly on the signal scenario, a common relative uncertainty value of 9\% on signal yield is used for all scenarios.

Uncertainties from limited Monte Carlo sample sizes, as well as uncertainties in the RS resonant shape due to the current knowledge of the electromagnetic energy scale of the calorimeter, the resolution of the detector and the pileup conditions were verified to have a negligible impact on the result. 
Theoretical uncertainties due to the PDFs and due to uncomputed higher perturbative orders were considered. 
The uncertainty in the PDFs translates into an uncertainty of $^{+10}_{-5}\,\%$ in the signal yields predicted by the ADD models. In the RS scenario the PDF uncertainty in the signal yield is between $6\,\%$ and $25\,\%$ for $m_{G}$ values between 0.5~TeV and 2.5~TeV.
As detailed in Ref.~\cite{ATLAS:2011ab}, an uncertainty of $\pm$0.1 is considered on the NLO K-factor value. The theory uncertainties are not included in the limit calculation, but their size is indicated in Figs.~\ref{fig:limits_xsec} and~\ref{fig:fiducial}, where graphical representations of the limits discussed in Section~\ref{sec:res} are shown.

\section{Results and interpretation}
\label{sec:res} 
A comparison of the observed invariant mass spectrum of diphoton events and the background expectation is shown 
in Fig.~\ref{fig:databkg}, along with the statistical significance of the bin-by-bin difference~\cite{significance} between data and the expected background. 
This significance is plotted as positive (negative) for bins 
with an excess (deficit) of data. Also shown are the expected signals
for two examples of RS and ADD models. The geometrical acceptance for the RS scenario with $m_{G}$ = 1.5 TeV and \coupling \ = 0.1 shown in Fig.~\ref{fig:databkg} is 88 \% and the selection efficiency for events within the acceptance is (54$\pm$5)\%. Both Fig.~\ref{fig:databkg}
and Table~\ref{tab:bkgestimation} show that, over the entire $m_{\gamma \gamma}$ range, the data are in good agreement with 
the expectations from SM backgrounds. 
The {\sc BumpHunter} algorithm~\cite{bumphunter} is used to further quantify the level of agreement between data and the SM background expectation.
The {\sc BumpHunter} performs a scan of the mass spectrum in windows 
of progressively increasing width. It identifies the window with the
most significant excess of data events over the background expectation
anywhere in a predefined search region. In this analysis, the binned
mass spectrum of Fig.~\ref{fig:databkg} is used, the search region is
defined as $142 < m_{\gamma\gamma} < 3000$~GeV and the window size is
allowed to vary from one bin to half the number of bins in the search
region\footnote{While the background is normalised to the data in the low-mass control region, the presence of a bump inside this region would still
be visible. This motivates the choice of the lower limit of the search region.}. The most significant excess is found in the region $1312 < m_{\gamma \gamma} < 1644$~GeV. The probability of observing, due to fluctuations in the background only, an excess that
is at least as significant as the one in data is $0.86$. This confirms
the absence of a significant excess in the mass spectrum.

\begin{figure}[h!]
\begin{center}
\includegraphics[width=\textwidth]{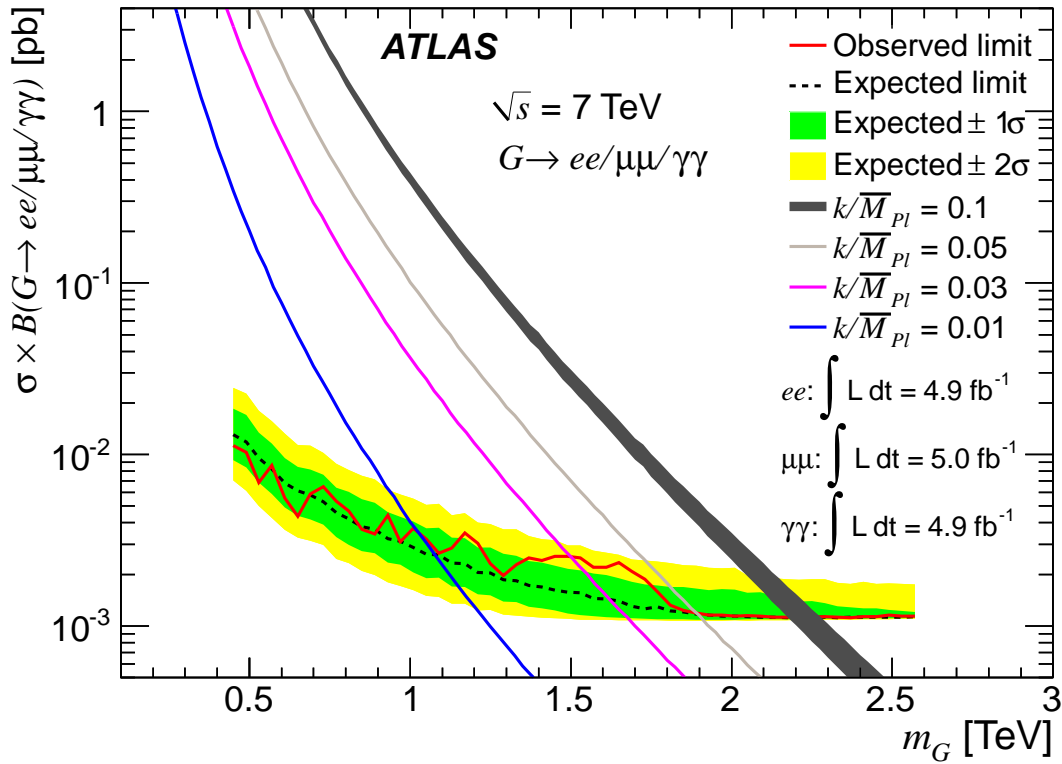}
\caption{Expected and observed $95\, \%$ CL limits from the
  combination of $G \rightarrow \gamma\gamma/ee/\mu\mu$ channels on
  $\sigma \times B$, the product of  the RS graviton production cross
  section and the branching ratio for graviton decay via $G
  \rightarrow \gamma\gamma/ee/\mu\mu$, as a function of the graviton
  mass. The --1$\sigma$ and --2$\sigma$ variations of the expected
  limit exhibit a tendency to be particularly close to the expected
  limit at large $m_{G}$. This behaviour is expected as signals with
  large $m_{G}$ would manifest themselves in regions of
  $m_{\gamma\gamma}$ where the SM background is small and the
  poissonian fluctuations around the mean expected background are
  highly asymmetric. The theory curves are obtained using the
  \Pythia\ generator, which implements the calculations from
  Ref.~\cite{TheoryCurve}. A K-factor of 1.75 is applied on top of
  these predictions to account for NLO corrections. The thickness of
  the theory curve for $\coupling = 0.1$ illustrates the theoretical
  uncertainties due to the PDFs expressed at 90\% CL. }
\label{fig:limits_xsec}
\end{center}
\end{figure}

\begin{figure}[h!]
\begin{center}
\includegraphics[width=\textwidth]{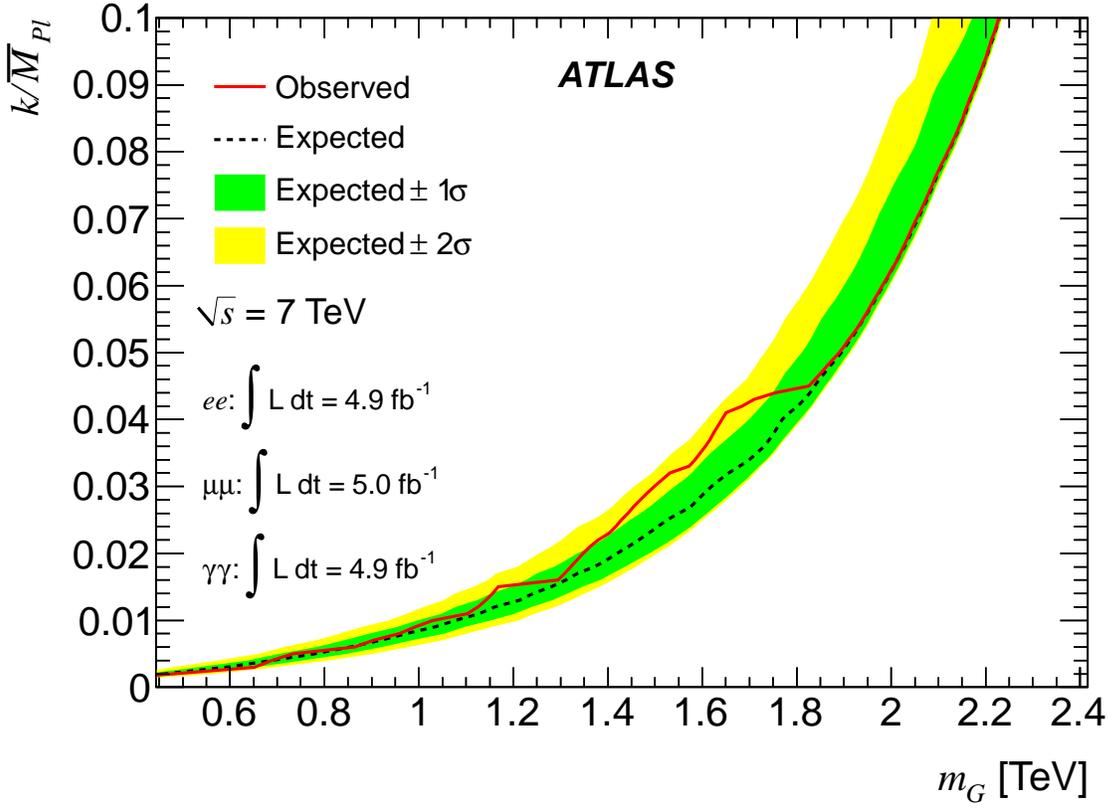}
\caption{The RS
  results interpreted in the plane of \coupling \ versus graviton
  mass. The region above the curve is excluded at $95\, \%$ CL.}
\label{fig:limits_couplingvsmg}
\end{center}
\end{figure}

Given the absence of evidence for a signal, $95\, \%$ CL upper limits were determined on the RS signal cross section.
The limits are computed using a Bayesian 
approach~\cite{bayes} assuming a flat prior for the RS cross section. The likelihood function is defined
as the product of the Poisson probabilities over all mass bins in the
search region defined as $m_{\gamma\gamma}>409$~\GeV. In each bin the
Poisson probability is evaluated for the observed number of data
events given the model prediction. The model prediction is the sum of the
expected background and the expected signal yield. The expected signal yield is a function of the graviton mass $m_{G}$ and the signal 
cross section times branching ratio $\sigma \times B(G \rightarrow \gamma \gamma)$. 
Systematic uncertainties are incorporated as nuisance parameters with Gaussian priors; the computation of the likelihood function includes integration over the variation in each of these. The RS model results can be combined with the results from the
dilepton channel~\cite{dilepton} to obtain limits on $\sigma \times B(G \rightarrow \gamma \gamma / ee / \mu\mu)$. Relative branching ratios of $B(G\rightarrow ee)/B(G\rightarrow \mu\mu)$ = 1 and $B(G\rightarrow \gamma\gamma)/B(G\rightarrow ee)$ = 2~\cite{TheoryCurve} are assumed for the purpose of this combination. Correlations between the systematic uncertainties 
for different channels are taken into account in the combination, as discussed in Ref.~\cite{ATLAS:2011ab}.
Specifically,  the systematic uncertainty in the QCD dijet background is treated as fully correlated across 
the $ee$ and $\gamma\gamma$ channels. The PDF and scale uncertainties are treated as correlated across all three channels, and affect the irreducible
background in the $\gamma \gamma$ channel as well as the Drell--Yan background in the $ee/\mu\mu$ channels.
The resulting limits are shown in Fig.~\ref{fig:limits_xsec}. 
The limit can be interpreted in the plane $(m_{G}, \coupling)$ using the theoretical dependence of the cross section on
these parameters (Fig.~\ref{fig:limits_couplingvsmg}). 
Alternatively, for a given value of \coupling, the limit can be translated into a 
limit on $m_{G}$. The limits on $m_{G}$ are summarised in
Table~\ref{tab:RSlimits}. Using a constant K-factor of 1.75, the $95\,
\%$ CL lower limit from the diphoton channel is 1.00~(2.06)~TeV for
\coupling= 0.01 (0.1), and the combined $95\, \%$ CL lower limit is
1.03~(2.23)~TeV for \coupling =0.01 (0.1).

\begin{table*}[bth]
\caption{$95\, \%$ CL lower limits on the mass $\left[{\rm TeV} \right]$ of the lightest RS
graviton, for various values of \coupling. The results are shown
for the diphoton channel alone  and for the combination of the
diphoton and dilepton channels, and also for both LO and NLO (K-factor~=~1.75) theory cross-section calculations.}
\begin{center}
\begin{tabular}{c|c|| c c c c} 
\hline
\hline
\multirow{3}{45pt}{\centering{K-factor value}}& \multirow{3}{60pt}{\centering{Channel(s) Used}}& \multicolumn{4}{c}{$95\, \%$ CL Observed (Expected) Limit [TeV]} \\ 
\cline{3-6}
                      &                                   &
\multicolumn{4}{c}{ \coupling \ value}  \\
                      &                                   & 0.01        & 0.03        & 0.05        & 0.1         \\
\hline\hline
\multirow{2}{*}{1 }   & \Ggg                              & 0.87 (0.88) & 1.31 (1.36) & 1.49 (1.60) & 1.91 (1.92) \\
                               & $G\to \gamma\gamma$/$ee$/$\mu\mu$ & 0.91 (0.95) & 1.39 (1.48) & 1.62 (1.75) & 2.10 (2.10) \\
\hline
\multirow{2}{*}{1.75} & \Ggg                              & 1.00 (0.98) & 1.37 (1.49) & 1.63 (1.73) & 2.06 (2.05) \\
                                 & $G\to \gamma\gamma$/$ee$/$\mu\mu$ & 1.03 (1.08) & 1.50 (1.63) & 1.89 (1.90) & 2.23 (2.23) \\
\hline
\hline
\end{tabular}
\label{tab:RSlimits}
\end{center}
\end{table*}  

A counting experiment is performed to set limits on the ADD model. Specifically, the number of diphoton events is counted in a search region above a given threshold in $m_{\gamma \gamma}$.
The mass threshold is chosen to optimise the expected limit on 
the difference in the diphoton cross section for $m_{\gamma \gamma} > 500$~GeV between the ADD model and the SM-only 
hypothesis. For the purpose of this optimisation, a specific implementation of the ADD model and specific values of the parameters have 
to be chosen. For $M_{S}$ = 2500 GeV in the GRW convention,  an optimal search region is obtained at 
$m_{\gamma \gamma} > 1217$~GeV. In the data, 4~events are observed in this search region, with a background expectation
of $2.32 \pm 0.37$ events. The expected and observed $95\, \%$~CL upper limits on the number of signal events above the SM
background in the search region are given in
Table~\ref{tab:limit_out}.

\begin{table}[h]
\caption{Expected and observed limits at $95\, \%$ CL on the number of
signal events in the search region at $m_{\gamma \gamma} > 1217$~GeV.}
\begin{center} 
\begin{tabular}{ccccc|c} 
\hline
\hline
\multicolumn{5}{c|}{Expected limit } & Observed limit\\ 
 $-2\sigma$ & $-1\sigma$ & mean & $+1\sigma$ & $+2\sigma$ &      \\  \hline \hline
3.08       & 3.08       & 5.18          & 5.96       & 8.53       & 7.21  \\ 
\hline
\hline
\end{tabular} 
\label{tab:limit_out}
\end{center} 
\end{table} 

\begin{figure}[htb]
\centering
\includegraphics[width=\textwidth]{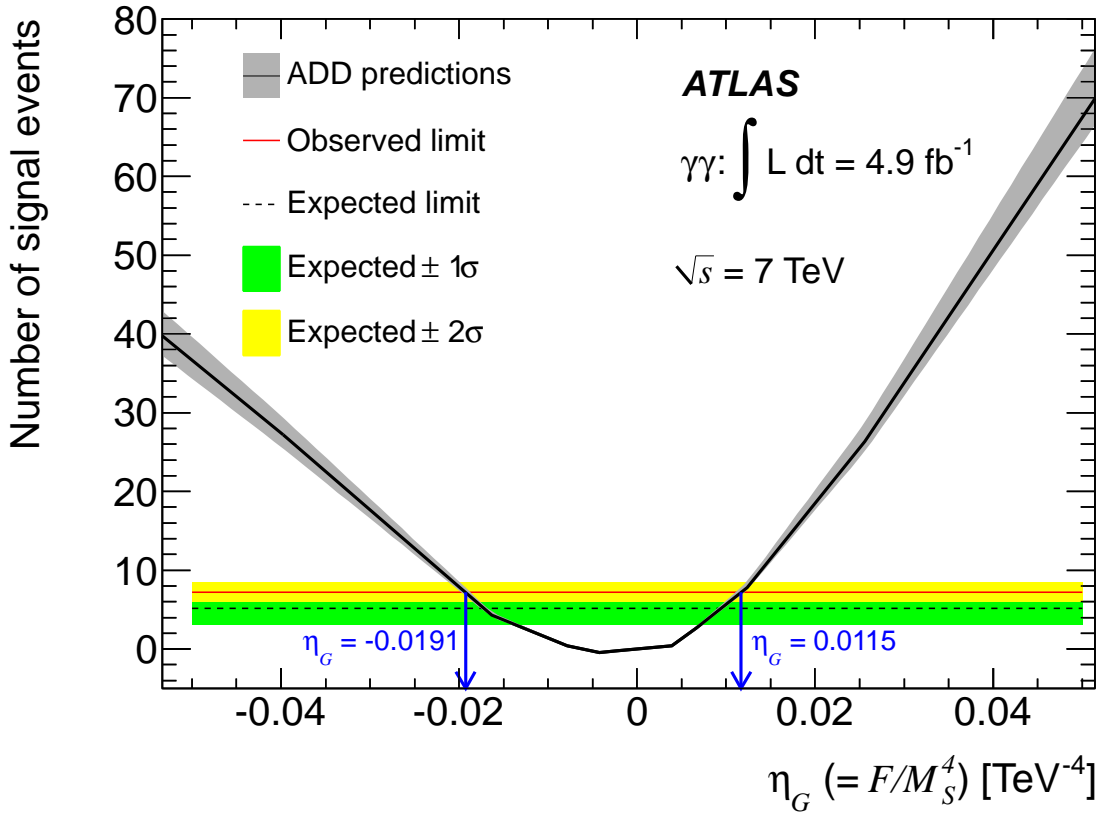}
\caption{Number of signal events as a function of $\eta_G$.
The solid horizontal line corresponds to the observed limit, the dashed line
corresponds to the expected limit and the bands to
the $\pm$1$\sigma$ and $\pm$2$\sigma$ uncertainty on the expected limit.
The black curve corresponds to MC predictions from various samples
without applying the K-factor of 1.70. The band around it illustrates the theoretical uncertainties due to the PDFs expressed at
90\% CL. The samples at positive (negative) $\eta_G$ have been simulated using the GRW (Hewett) formalism.
When the prediction is greater than the observed limit, the corresponding value for $\eta_G$ is excluded.}
\label{fig:fiducial}
\end{figure}

These limits on the event yield can be translated into limits 
on the parameter $\eta_G$ of the ADD model using a prediction of the expected excess of events over the SM-only background as
a function of $\eta_G$. This translation is shown, for one specific implementation of the ADD model, in Fig.~\ref{fig:fiducial}.

The limits\footnote{The limit in the HLZ convention with n=2 may appear weak compared to the corresponding limit reported in Ref.~\cite{CMSggnew}. When adopting the same search region as in Ref.~\cite{CMSggnew}, namely $m_{\gamma \gamma} > 901$~GeV instead of $m_{\gamma \gamma} > 1217$~GeV, our HLZ, n=2 limit for a K-factor=1 becomes $M_{s}$=3.57~TeV, which is more stringent thant the result obtained with our default search region.} for different ADD scenarios are summarised in Table~\ref{tab:limit_para}. Using a constant K-factor of 1.70, 
the $95\, \%$ CL upper (lower) limit on $\eta_G$ is $0.0085$ ($-0.0159$)~TeV$^{-4}$ for constructive (destructive) interference.

\begin{table*}[htb]
\footnotesize 
\caption{ Observed 95\% CL limits on the ADD model parameters $\eta_{G}$~$\left[{\rm TeV}^{-4} \right]$ and $M_{S}$~$\left[{\rm TeV} \right]$ for various ADD models.}
\begin{center}
\begin{tabular}{c|l||c|cc|cccccc} 
\hline
\hline
K-factor & ADD & GRW & \multicolumn{2}{c|}{Hewett} & \multicolumn{6}{c}{HLZ} \\
Value &  Parameter  &     & Neg. & Pos. & $n=2$ & $n = 3$ &$n = 4$ &$n = 5$ &$n = 6$ &$n = 7$    \\ \hline \hline
1         & $\eta_{G} $ & 0.0115 & --0.0191& 0.0115  & \multicolumn{6}{c}{0.0115}  \\ 
            & $M_{s}$    & 3.05  & 2.40 & 2.73   & 3.11 &  3.63 & 3.05 & 2.76 & 2.57 & 2.43  \\\hline
1.70    & $\eta_{G} $ & 0.0085 & --0.0159 & 0.0085  & \multicolumn{5}{c}{0.0085}   \\ 
             & $M_{s}$     & 3.29  & 2.52 & 2.94  & 3.55 & 3.92 & 3.29 & 2.98 & 2.77 & 2.62 \\
\hline
\hline
\end{tabular}
\label{tab:limit_para}
\end{center}
\end{table*}

\section{Summary}

A search for evidence of extra dimensions has been performed in the diphoton channel, based on the full 2011 dataset collected by the ATLAS 
detector at the LHC ($4.9$~fb$^{-1}$ of proton--proton collisions at $\sqrt{s} =
7$~TeV). In the ADD scenario, lower limits of between 2.52 and 3.92~TeV at $95\,\%$ CL are set on the $M_S$~scale, 
depending on the number of extra dimensions and the theoretical
formalism used. The combination with the dilepton channel presented in Ref.~\cite{PhysRevD.87.015010} further tightened these limits to 2.6--4.2~TeV.

In the RS scenario, a lower limit of 1.00 (2.06)~TeV at $95\,\%$ CL is set on the mass of the 
lightest RS graviton, for RS couplings of $\coupling = 0.01 \, (0.1)$. These new limits are a significant improvement over the previous best limit in the diphoton channel of 0.80 (1.85) TeV for $\coupling = 0.01 \, (0.1)$~\cite{ATLAS:2011ab}. The combination with the latest ATLAS dilepton searches~\cite{dilepton} yields an improvement by a factor of 2-3 in the cross-section limits compared to Ref.~\cite{ATLAS:2011ab} and further tightens the RS graviton mass limits to 1.03 (2.23) TeV for $\coupling = 0.01 \, (0.1)$.

\vspace{0.2cm}


\section{Acknowledgements}

We thank CERN for the very successful operation of the LHC, as well as the
support staff from our institutions without whom ATLAS could not be
operated efficiently.

We acknowledge the support of ANPCyT, Argentina; YerPhI, Armenia; ARC,
Australia; BMWF and FWF, Austria; ANAS, Azerbaijan; SSTC, Belarus; CNPq and FAPESP,
Brazil; NSERC, NRC and CFI, Canada; CERN; CONICYT, Chile; CAS, MOST and NSFC,
China; COLCIENCIAS, Colombia; MSMT CR, MPO CR and VSC CR, Czech Republic;
DNRF, DNSRC and Lundbeck Foundation, Denmark; EPLANET and ERC, European Union;
IN2P3-CNRS, CEA-DSM/IRFU, France; GNSF, Georgia; BMBF, DFG, HGF, MPG and AvH
Foundation, Germany; GSRT, Greece; ISF, MINERVA, GIF, DIP and Benoziyo Center,
Israel; INFN, Italy; MEXT and JSPS, Japan; CNRST, Morocco; FOM and NWO,
Netherlands; BRF and RCN, Norway; MNiSW, Poland; GRICES and FCT, Portugal; MERYS
(MECTS), Romania; MES of Russia and ROSATOM, Russian Federation; JINR; MSTD,
Serbia; MSSR, Slovakia; ARRS and MVZT, Slovenia; DST/NRF, South Africa;
MICINN, Spain; SRC and Wallenberg Foundation, Sweden; SER, SNSF and Cantons of
Bern and Geneva, Switzerland; NSC, Taiwan; TAEK, Turkey; STFC, the Royal
Society and Leverhulme Trust, United Kingdom; DOE and NSF, United States of
America.

The crucial computing support from all WLCG partners is acknowledged
gratefully, in particular from CERN and the ATLAS Tier-1 facilities at
TRIUMF (Canada), NDGF (Denmark, Norway, Sweden), CC-IN2P3 (France),
KIT/GridKA (Germany), INFN-CNAF (Italy), NL-T1 (Netherlands), PIC (Spain),
ASGC (Taiwan), RAL (UK) and BNL (USA) and in the Tier-2 facilities
worldwide.

\section*{References}
\bibliographystyle{atlasBibStyleWithTitle}
\bibliography{Paper_diphoton}

\onecolumn 
\clearpage
\begin{flushleft}
{\Large The ATLAS Collaboration}

\bigskip

G.~Aad$^{\rm 48}$,
T.~Abajyan$^{\rm 21}$,
B.~Abbott$^{\rm 111}$,
J.~Abdallah$^{\rm 12}$,
S.~Abdel~Khalek$^{\rm 115}$,
A.A.~Abdelalim$^{\rm 49}$,
O.~Abdinov$^{\rm 11}$,
R.~Aben$^{\rm 105}$,
B.~Abi$^{\rm 112}$,
M.~Abolins$^{\rm 88}$
% ...
\bigskip
\\
% ...
$^{11}$ Institute of Physics, Azerbaijan Academy of Sciences, Baku, Azerbaijan\\
$^{12}$ Institut de F{\'\i}sica d'Altes Energies and Departament de F{\'\i}sica de la Universitat Aut{\`o}noma de Barcelona and ICREA, Barcelona, Spain\\
% ...
$^{21}$ Physikalisches Institut, University of Bonn, Bonn, Germany\\
% ...
$^{48}$ Fakult{\"a}t f{\"u}r Mathematik und Physik, Albert-Ludwigs-Universit{\"a}t, Freiburg, Germany\\
$^{49}$ Section de Physique, Universit{\'e} de Gen{\`e}ve, Geneva, Switzerland\\
% ...
$^{88}$ Department of Physics and Astronomy, Michigan State University, East Lansing MI, United States of America\\
% ...
$^{105}$ Nikhef National Institute for Subatomic Physics and University of Amsterdam, Amsterdam, Netherlands\\
% ...
$^{111}$ Homer L. Dodge Department of Physics and Astronomy, University of Oklahoma, Norman OK, United States of America\\
$^{112}$ Department of Physics, Oklahoma State University, Stillwater OK, United States of America\\
% ...
$^{115}$ LAL, Universit{\'e} Paris-Sud and CNRS/IN2P3, Orsay, France\\
% ...
\end{flushleft}


\onecolumn
\clearpage

\end{document}